\documentclass[10pt,prl,aps,twocolumn,showpacs]{revtex4}
\usepackage{graphicx}
\usepackage{amsmath}

\begin{document}

\title{Deconfinement and criticality in extended two-dimensional dimer models}

\author{Anders W. Sandvik}
\affiliation{Department of Physics, {\AA}bo Akademi University,
Porthansgatan 3, FIN-20500 Turku, Finland}

\begin{abstract}
A square-lattice hard-core dimer model with links extending beyond 
nearest-neighbors is studied using a directed-loop Monte Carlo method. An 
arbitrarily small fraction of next-nearest-neighbor dimers is found to 
cause deconfinement, whereas a critical state with $r^{-2}$ distance 
dependence of the dimer-dimer correlations persists in the presence of 
longer dimers preserving the bipartite graph structure. However, the 
critical confinement exponent governing the correlation of two test 
monomers is non-universal. Implications for resonating-valence-bond 
states are discussed.
\end{abstract}

\date{December 3, 2003}

\pacs{74.20.Mn, 75.10.-w, 05.10.Ln, 05.50.+q}

\maketitle

Dimer models have a long history in classical statistical physics
\cite{fow37a,fis61a,kas61a,fis63a}. More recently, they have also emerged as 
central models in modern theories of strongly correlated quantum matter, e.g.,
high-temperature cuprate superconductors and frustrated antiferromagnets 
\cite{kiv87a,rok88a,fra91a,sen00a,mis02a,sac03a,ard03a}. The dimers then 
represent singlet-forming electron pairs. In order to model the quantum
fluctuations of the dimers and realize a short-range version of Anderson's 
resonating valence bond (RVB) state \cite{and73a,and87a}, Kivelson, Rokhsar, 
and Sethna introduced a Hamiltonian with a term flipping (resonating) pairs 
of parallel nearest-neighbor dimers on the two-dimensional (2D) square lattice
\cite{kiv87a}. The purely classical dimer model retains it relevance also 
here: It was shown that the equal-weight sum over all dimer configurations
is the ground state of the Hamiltonian when the resonance strength $-k$ equals
the potential energy cost $v$ of each resonating pair of dimers (the RK point)
\cite{rok88a}. This state is critical; the dimer-dimer correlations decay 
with distance as $r^{-2}$ and two inserted test monomers are correlated with 
each other as $r^{-1/2}$ \cite{fis63a}. As it turned out, away from the RK 
point the dimers form long-range order and the monomers are exponentially 
confined \cite{sac89a,fra91a,leu96a}. Hence this system does not give rise to 
the desired RVB state with  no broken lattice symmetries and deconfined 
monomers (corresponding to spin-charge  separation \cite{and87a}). Moessner 
and Sondhi recently showed that a true extended RVB phase {\it does} appear in
the quantum dimer model on the triangular lattice \cite{moe01a}. Following this 
insight, a large body of work has been carried out in order to characterize 
classical and quantum dimer models on various lattices
\cite{moe01b,mis02a,fen02a,hus03a,moe03a,fra03a}. Moreover, there are 
currently intense activities in gauge theories related to quantum dimer models 
\cite{sen00a,moe01b,mis02a,sac03a,ard03a,vis03a}.

To date, research on dimer models has focused mainly on planar lattices, 
i.e., ones that have no intersecting links. This
class of models can be analytically solved (in the form of Pfaffians) 
with the aid of a theorem by Kastelyn \cite{kas61a}, and thus the quantum 
ground states at the corresponding RK points are characterized as well.
It has often been stated that the inclusion of valence bonds (dimers) 
extending further than between nearest-neighbor sites will not change the 
physics provided that the probability of longer bonds decreases sufficiently 
rapidly \cite{rok88a,moe01b}. However, the fact that there are qualitative 
differences between bipartite and nonbipartite lattices, e.g., the square and 
triangular cases mentioned above, does raise the question of potentially 
important effects of short bonds connecting two sites on the  same sublattice 
of the square lattice. Such bonds will inevitably appear in realistic systems 
away from the limiting cases \cite{kiv87a,cha89a,rea89a} represented by 
the nearest-neighbor ($N_1$) dimer models. Introducing next-nearest-neighbor 
($N_2$) links along one of the diagonals makes the square lattice equivalent 
to the triangular one. It has already been shown that 
an arbitrarily small fraction of such diagonal dimers destroys the 
critical $N_1$ square-lattice state and leads to deconfinement 
\cite{fen02a,ard03a}, as in the isotropic triangular lattice \cite{moe01a}. 
The model with links along both diagonal directions, i.e., the full 2D square 
lattice with $N_1$ and $N_2$ bonds, is not solvable by Kastelyn's theorem 
\cite{kas61a}. Intuitively, one might suspect that the critical state is 
immediately destroyed in this case as well, but no calculations have been 
carried out thus far. One might also speculate that introducing longer bonds 
between the two sublattices does not lead to deconfinement, but there are no 
results available to back this up.

In this Letter, two extended dimer models on the square lattice are 
studied---the nonbipartite model with $N_1$ and $N_2$ dimers as well as the 
bipartite lattice with $N_1$ and $N_4$ (fourth-nearest-neighbors, of which
there are eight per site). An efficient 
directed-loop \cite{syl02a} Monte Carlo algorithm is used to 
sample the full space of hard-core dimer configurations, with fugacities $w_i$ 
assigned to the different dimer types. The results confirm that a low, most 
likely infinitesimal, concentration of $N_2$ dimers leads to deconfinement, 
and that the presence of $N_4$ dimers does not. However, while the dominant
dimer correlations are $\sim$$1/r^2$ also when $w_4 > 0$, those correlations
involve $N_4$ dimers; the $N_1$ correlations decay as a higher power of
$1/r$. The structure in wave-vector space also becomes much more 
complicated. Furthermore, {\it the critical confinement exponent governing 
the monomer correlation function is nonuniversal}, changing continuously 
from $-1/2$ in the pure $N_1$ case to $-1/9$ (to high numerical accuracy) 
for the pure $N_4$ model.

{\it Directed-loop algorithm}.---This algorithm is an adaptation of a 
quantum Monte Carlo method \cite{syl02a}, with the same name, in which updates 
of the system degrees of freedom are carried out along a self-intersecting path
at the endpoints of which there are defects not allowed in the configuration
space contributing to the partition function. When the two defects meet
they annihilate, the loop closes, and a new allowed configuration is
created. The conditions for detailed balance in the process of 
stochastically moving one of the defects are expressed as a coupled set of 
{\it directed-loop equations} \cite{syl02a}. The applicability of this 
scheme to dimer models was first realized by Adams and Chandrasekharan 
\cite{ada03a} (an alternative cluster algorithm was constructed in 
Ref.~\cite{kra03a}). Here an algorithm for the multi-length dimer problem 
will be presented; a simplifying representation of the dimer configurations 
will also be introduced. The algorithm will be described only for the case 
of the $N_1$-$N_2$ model, but the scheme applies directly to any range of 
the links.

\begin{figure}
\includegraphics[width=7.5cm]{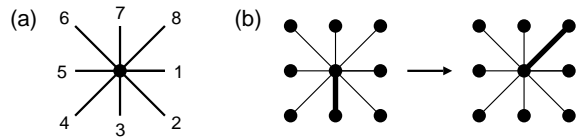}
\caption{(a) Labeling of the links of the $N_1$-$N_2$ dimer model.
(b) A step in the directed-loop update, in which the entrance to the
vertex is at site 3 and the exit is at site 8. The thick bond indicates
the location of the dimer.}
\label{fig1}
\end{figure}

The links connected to a given site are labeled as shown in Fig.~\ref{fig1}(a).
For a dimer configuration on a periodic $L\times L$ lattice, each site is 
numbered according to the type of dimer it is connected to. For any given site,
the other member of the same dimer can then be easily found. The central object
in the directed-loop algorithm is a {\it vertex}, which in this case consists 
of a site and all its surrounding sites to which it can be coupled by a dimer.
The state of the vertex is the number ($1$-$8$) assigned to the central site.
A step in the directed-loop algorithm is illustrated in Fig.~\ref{fig1}(b). 
The vertex is entered through the dimer, and one of the 
seven other surrounding sites is chosen as the exit. The dimer is then 
flipped from the entrance to the exit, the central vertex site remaining 
connected to it. The exit site already has another dimer connected to it, and 
this site now becomes the entrance in the next step of the algorithm. In each
step a defect is hence moved one link ahead, leaving behind a healed dimer 
state. The first entrance site is chosen at random and the corresponding 
defect remains stationary. It is annihilated when the moving defect reaches 
it, whence a new allowed dimer configuration has been generated. In the 
present case the defects are monomers, and the intermediate two-monomer 
configurations can thus be used to measure the correlations between two 
inserted test monomers.

The key to an efficient algorithm of this kind is that the probabilities for 
random selection of the seven possible exit sites can be chosen in such a way
that detailed balance is satisfied without any further accept/reject criterion.
Each step then moves one dimer, and a full loop can accomplish very 
significant changes to the dimer configuration. The directed-loop
equations \cite{syl02a} give the conditions for detailed balance in terms of 
weights $a_{jk}$  for the processes in which a vertex in state $j$ is entered 
at site $j$ and exited at $k$ (transforming the vertex into state $k$). 
The actual probabilities $P_{jk}=a_{jk}/w_j$, 
which implies $\sum_k a_{jk}=w_j$. Detailed balance is satisfied if $a_{jk}=
a_{kj}$. In the $N_1$-$N_2$ model, the vertices can be classified as even 
($e$) or odd ($o$) according to the numbering of Fig.~\ref{fig1}(a); there are 
then four weights: $a_{ee},a_{oo},a_{eo},a_{oe}$. In principle, one can include
``bounce'' processes where $j=k$, but in the present case they can be excluded.
Including only the seven no-bounce exits, the directed-loop equations reduce to
\begin{subequations}
\begin{eqnarray}
w_1 & = & 3a_{oo}+4a_{oe}, \\
w_2 & = & 3a_{ee}+4a_{eo}, \\
a_{eo} & = & a_{oe}.
\end{eqnarray}
\end{subequations}
This system is underdetermined and has an infinite number of postive-definite
solutions. Here the following solution will be used: For $w_1 \ge w_2$,
\begin{subequations}
\begin{eqnarray}
a_{ee} & = & a_{oe}=a_{eo}=w_2/7, \\
a_{oo} & = & (w_1 - 4w_2/7)/3,
\end{eqnarray}
\end{subequations}
while for $w_2 \ge w_1$,
\begin{subequations}
\begin{eqnarray}
a_{oo} & = & a_{oe}=a_{eo}=w_1/7, \\
a_{ee} & = & (w_2 - 4w_1/7)/3.
\end{eqnarray}
\end{subequations}
There is no guarantee that this is the best solution, but the resulting 
algorithm performs very well and allowed for studies of lattices 
with $\sim$$10^6$ dimers. The fugacity of the $N_1$-bonds is set to 
unity (except in the case of the pure $N_2$- and $N_4$-models). The program 
was tested using known results for the pure $N_1$ case \cite{fis63a} and 
by comparing with local Metropolis simulations for small lattices.

\begin{figure}
\includegraphics[width=6.5cm]{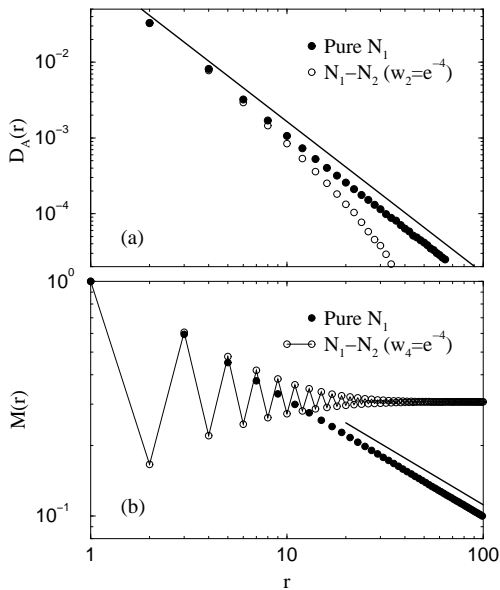}
\caption{Dimer (a) and monomer (b) correlations
along the direction $(r,0)$ for the $N_1$-$N_2$ model at $w_2=e^{-4}$, compared
with those of the pure $N_1$ model. The solid lines show the agreement with 
the known power-laws \cite{fis63a} for the $N_1$ model. Note that for the pure
$N_1$ model $M(r)=0$ for even $r$. The results were obtained using $L=1024$ 
lattices.}
\label{fig2}
\end{figure}

{\it Results.}---Dimer-dimer correlations in the full close-packed system and 
monomer-monomer correlations in the system with two test monomers will be 
discussed. The monomer correlations $M(\mathbf{r})$ were obtained by 
accumulating the distances between the stationary and the moving monomer in 
the directed-loop update. As has become customary \cite{kra03a}, the 
normalization $M(r=1)=1$. Several types of dimer-dimer correlations can 
be defined. Here $D_\Sigma(\mathbf{r})$ will be defined in the following way: 
If site $i$ is connected to a dimer in the set $\Sigma$, a variable $s(i)=1$,
otherwise $s(i)=0$. The correlation function is then $D_\Sigma(\mathbf{r}_{ij})
=\langle s(i)s(j)\rangle$. Results will be presented 
for cases where $\Sigma$ contains a single dimer or half of the dimers of one 
type. For the $N_1$-$N_2$ model, the correlation functions $D_1$, $D_2$, 
$D_A$, and $D_B$ are thus defined corresponding to the sets $\{ 1\}$, $\{ 2\}$,
$A=\{ 1,3\}$, and $B=\{ 2,4\}$. Analogous definitions are used
for correlations $D_1$, $D_A$ of $N_1$ dimers and $D_4$, $D_D$ of $N_4$
dimers in the $N_1$-$N_4$ model. 

In Fig.~\ref{fig2}(a), dimer correlations $D_A$ for the nonbipartite 
$N_1$-$N_2$ model with $w_2 = {\rm e^{-4}}$ (corresponding to a concentration 
$p_2 \approx 0.3\%$ of $N_2$ dimers) are compared with those of the pure $N_1$
model. There is a clear deviation from the $r^{-2}$ decay, showing that the 
$N_1$-$N_2$ model is not critical at this very low concentration of $N_2$ 
dimers. As shown in Fig.~\ref{fig2}(b), the test monomer correlation approaches
a nonzero constant, i.e., the system is deconfined in contrast to the 
critically confined $N_1$ model. The very significant changes seen already
at a very low concentration of $N_2$ dimers suggest that an arbitrarily
small concentration indeed causes deconfinement. 

\begin{figure}
\includegraphics[clip,width=7cm]{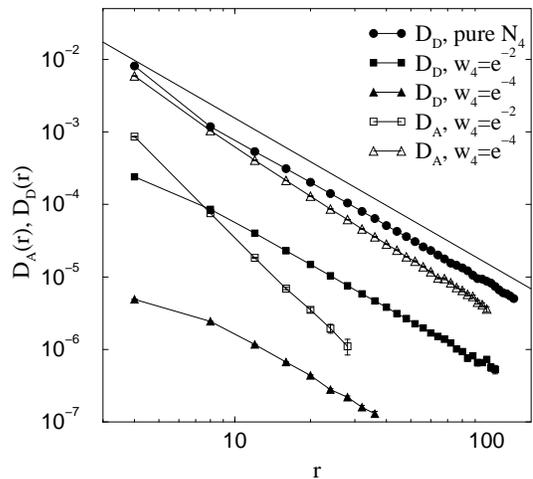}
\caption{Various dimer correlations along the direction $(r,0)$ in the 
$N_1$-$N_4$ model (on $L=1024$ lattices). The fugacities $w_4={\rm e}^{-4}$
and ${\rm e}^{-2}$ correspond to concentrations $p_4\approx 0.065$ and $0.17$, 
respectively. The solid line shows the asymptotic form $r^{-2}$.}
\label{fig3}
\end{figure}

\begin{figure}
\includegraphics[width=8.4cm]{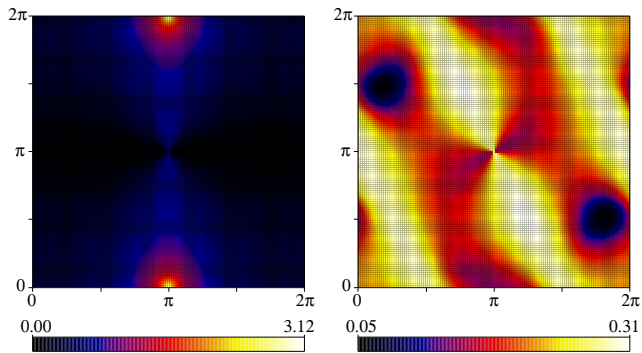}
\caption{Fourier transform of the dimer-dimer correlation $D_1$ of the $N_1$ 
model (left) and $D_4$ of the $N_4$ model (right) calculated on $L=128$ 
lattices.}
\label{fig4}
\end{figure}

Turning now to the bipartite $N_1$-$N_4$ model, its dimer correlations
$D_A(r)$ ($N_1$ dimers) and  $D_D(r)$ ($N_4$ dimers) are compared with
$D_A(r)$ of the pure $N_1$ model in Fig.~\ref{fig3}. Surprisingly, the two 
correlation functions decay with different power-laws: $D_D(r)$ apparently 
always follows the same form $r^{-2}$ as $D_A(r)$ of the pure $N_1$ model, 
whereas $D_A(r)$ decays faster once $w_4 > 0$. In Fig.~\ref{fig4} the Fourier 
transforms of the correlation functions involving a single type of dimer are 
shown for the pure $N_1$ and $N_4$ models. For the $N_1$ model, the dominant 
correlations are at $\mathbf{q}=(\pi,0)$. A logarithmic divergence of the 
peak height with the system size can easily be observed (not shown here). The 
$N_4$ model exhibits more complicated correlations, with very broad peaks that 
show almost no size dependence up to the largest size ($L=2048$) that was 
studied. Nevertheless, a logarithmic divergence should eventually occur 
here as well, considering the $r^{-2}$ real-space correlations. The details of
the nature of these critical correlations remain to be elucidated. In the 
$N_1$-$N_4$ model, the wave-vector structures shown in Fig.~\ref{fig4} change
with $w_4$. Most notably, the $D_1$ peak at $(\pi,0)$ is suppressed as the 
fraction of $N_4$ dimers increases, and a nondivergent structure at 
$(\pi,\pi)$ emerges. 

\begin{figure}
\includegraphics[width=6.75cm]{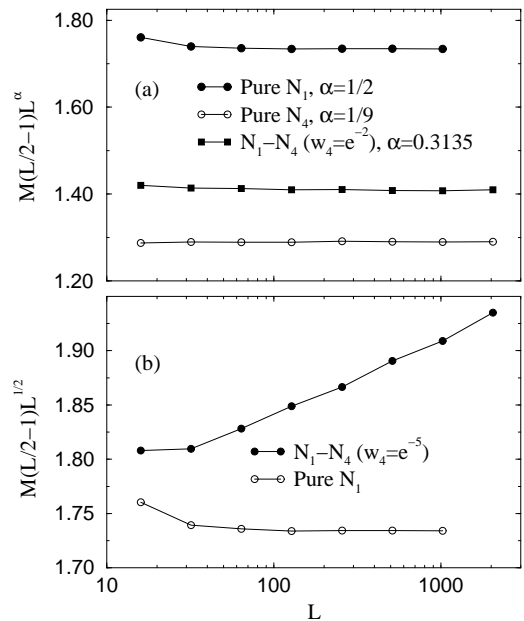}
\caption{Finite-size scaling of monomer correlations in the $N_1$-$N_4$
model. (a) The best scaling for three different cases. (b) Failure of a 
scaling with the pure $N_1$ exponent $\alpha=1/2$ for the $N_1$-$N_4$ model 
at low fugacity $w_4$.}
\label{fig5}
\end{figure}

An unchanged dominant critical dimer exponent might suggest that the monomer 
correlations should also remain of the pure $N_1$ form $r^{-1/2}$. This is 
not the case, however. Results for $M(r)$ at $r=L/2-1$ are shown multiplied by 
$L^\alpha$ in Fig.~\ref{fig5}(a). Here $\alpha$ is adjusted to give a flat 
$L$ dependence, and hence the critical form $M(r)\sim r^{-\alpha}$ is 
extracted. The known $\alpha=1/2$ is used for the pure $N_1$ model. For 
the pure $N_4$ model the exponent is consistent with $\alpha =1/9$ (to
an accuracy of $1\%$). To show that the exponent changes from 
$1/2$ already at a low concentration of $N_4$ bonds, results for 
$w_4={\rm e}^{-5}$ ($p_4\approx 0.9\%$) are scaled with $\alpha =1/2$ in
Fig.~\ref{fig5}(b). This scaling clearly fails, and is instead consistent 
with $\alpha \approx 0.485$.

{\it Conclusions.}---The results obtained here demonstrate that the 2D 
square-lattice dimer model becomes deconfined when a very low concentration 
(likely infinitesimal) of non-bipartite (next-nearest-neighbor) dimers are 
introduced. The system remains critically confined in the presence of longer 
bipartite (fourth-nearest-neighbor) dimers, but the corresponding two-monomer 
exponent is nonuniversal. In contrast, the dimer-dimer correlation exponent 
appears to remain unchanged, although the nature of the dominant correlations 
changes.

These findings are relevant to quantum dimer
models as well. Various resonance terms can be introduced, and corresponding
RK points can then be demonstrated in the same way as has been done for other
models \cite{rok88a,ard03a,hen03a}. Hence, it is clear that an RVB 
state can be realized on the square lattice once next-nearest-neighbor dimers 
are allowed. On the other hand, the emergence of new structure in the 
dominant dimer-dimer correlations on the bipartite graph including 
fourth-nearest-neighbor links indicates that phase transitions between
different long-range ordered valence-bond-solid states can be realized. 
Such quantum order-order transitions have recently been discussed by 
Vishwanath, Balents, and Senthil \cite{vis03a}. Extended quantum dimer models 
on the square lattice should thus have very rich phase diagrams and may also 
be relevant in the context of the cuprates.

This work was supported by the Academy of Finland, Project No.~26175. The
calculations were carried out on the Condor system at the University of
Wisconsin-Madison.

\end{document}